\documentclass[10pt, conference]{IEEEtran}
\usepackage{cite}
\usepackage{hyperref}

\ifCLASSINFOpdf
\else
\fi
\usepackage{graphicx}
\usepackage{subfigure}

\usepackage[cmex10]{amsmath}
\usepackage{amssymb}
\usepackage{amsthm}

\newtheorem{theorem}{Theorem}
\newtheorem{lemma}{Lemma}
\usepackage{algorithmic}
\usepackage[linesnumbered, ruled, vlined]{algorithm2e}

\usepackage{tabularx}

\usepackage{url}

\usepackage[T1]{fontenc}

\begin{document}
%
\title{Minimum Latency Broadcast Scheduling in Single-Radio Multi-Channel Wireless Ad-Hoc Networks}

\author{Lizhao You$^{\dagger}$, Zimu Yuan$^{\S}$, Bin Tang$^{\dagger}$, Guihai Chen$^{\dagger}$\\
$^{\dagger}$State Key Laboratory for Novel Software Technology, Nanjing University, China \\
$^{\S}$Institute of Computing Technology, CAS, China\\
Email: lzyou@smail.nju.edu.cn, yuanzimu@ict.ac.cn, tb@dislab.nju.edu.cn, gchen@nju.edu.cn\\
}


%


\maketitle

\begin{abstract}
We study the minimum latency broadcast scheduling (MLBS)
problem in Single-Radio Multi-Channel (SR-MC) wireless ad-hoc
networks (WANETs), which are modeled by Unit Disk Graphs. Nodes
with this capability have their fixed reception channels, but
can switch their transmission channels to communicate with
their neighbors. The single-radio and multi-channel model
prevents existing algorithms for single-channel networks
achieving good performance. First, the common assumption
that one transmission reaches all the neighboring nodes does
not hold naturally. Second, the multi-channel dimension
provides new opportunities to schedule the broadcast
transmissions in parallel. We show MLBS problem in SR-MC WANETs
is NP-hard, and present a benchmark algorithm: Basic
Transmission Scheduling (BTS), which has approximation ratio of
$4k+12$. Here $k$ is the number of orthogonal channels in SR-MC
WANETs. Then we propose an Enhanced Transmission Scheduling
(ETS) algorithm, improving the approximation ratio to $k+23$.
Simulation results show that ETS achieves better
performance over BTS, and the performance of ETS
approaches the lower bound.
\end{abstract}



%
\IEEEpeerreviewmaketitle


\section{Introduction}
Single-Radio Multi-Channel (SR-MC) wireless ad hoc networks (WANETs) have gained significant attentions in the past few years because of their great promise of low cost, high throughput and spectral efficiency. By using multiple orthogonal channels in single radio, we can enhance spatial reuse \cite{MMSN}, alleviate jamming attack \cite{Jamming}, and enable dynamic access to the scarce spectrum resource \cite{CRAHNs}. Several typical multi-channel MACs \cite{xRDT, CCR-MAC, Quorum} are proposed to fully utilize the single-radio multi-channel capability.

Broadcast is a fundamental operation in wireless networks for
routing discovery, information dissemination, and so on. Here
we focus on the minimum latency broadcast scheduling operation,
where broadcast latency is defined as the end-to-end latency by
which all nodes in the network receive the broadcast message
from source node. Such concern is very important in various
applications such as military communications, disaster relief
and rescue operations.

For some SR-MC networks \cite{CCR-MAC}, broadcast packets can be delivered in a dedicated control channel (DCC). However, the DCC can be a bottleneck, vulnerable to jamming attack \cite{Jamming}, and even unavailable \cite{CRAHNs}. Therefore, in this paper, we consider SR-MC WANETs without the DCC. Given no DCC, the solutions to MLBS problem in single-channel WANETs \cite{UDG, info07} cannot work directly, because single transmission cannot reach all the neighboring nodes if the radios of neighboring nodes are tuned to different channels. In other words, it may cost multiple transmissions to deliver a message to all its neighbors. This property is similar to the \emph{partial broadcast property} in duty-cycled WANETs \cite{ICC09, GC11}. However, parallel transmissions can still happen in different nodes within several channels at the same time (we refer this property as \emph{multi-partial broadcast property}), which is not allowed in duty-cycled networks. Hence, solutions from duty-cycled networks cannot achieve best performance, and can be further optimized. On the other hand, compared with solutions for MR-MC WANETs \cite{MR-MC}, single-radio mode does not allow a node to transmit simultaneously in several channels, resulting in less parallel transmission opportunities. Therefore, \emph{multi-partial broadcast property} brings a new challenge for designing efficient, collision-free broadcast protocols.

In this paper, we investigate the minimum latency broadcast
scheduling (MLBS) problem in SR-MC WANETs. We first show such
problem is NP-Hard, and then design efficient algorithms with
performance guarantees. To solve the problem, we construct a
Shortest-Path Tree (SPT), and schedule the transmissions layer
by layer. By utilizing the \emph{multi-partial broadcast
property}, we can schedule the cross-layer and same-layer
transmissions with polynomial-time complexity. Our main
contributions are summarized as follows:
\begin{itemize}
  \item We show that a Basic Transmission Scheduling (BTS) algorithm with approximation ratio of 4k+12 can be obtained by modifying existing approaches properly, where $k$ is the number of available orthogonal channels.
  \item We present an Enhanced Transmission Scheduling (ETS) algorithm by utilizing the parallel transmission opportunities, which has an improved approximation ratio of $k+23$. The performance is evaluated through extensive simulations.
\end{itemize}

The rest of the paper is organized as follows. Section \ref{rw}
gives the related work. Network model and problem statement are
presented in Section \ref{pre}. We first propose BTS in Section
\ref{bts}, and give ETS in Section \ref{ets}. Then we validate
our result by simulations in Section \ref{eval}. Section
\ref{con} concludes our paper.

\section{Related Works} \label{rw}
Channel assignment in SR-MC WANETs is the most related works to
ours. In general, there are three kinds of channel assignment
approaches: \emph{fixed}, \emph{semi-dynamic} and
\emph{dynamic}. In fixed channel assignment method, nodes are
assigned fixed channels for permanent use, and radios do not
change the operating frequency. In \emph{semi-dynamic}
approaches, though the assigned reception channel is fixed,
nodes can still change their transmission channel to
communicate with neighbors that have different reception
channels. In dynamic approaches, nodes are not assigned static
channels, and can switch their channel dynamically according to
a pre-defined rule, e.g., quorum sequences \cite{Quorum}.
Moreover, some works consider channel assignment and other
problems jointly, e.g., minimizing interference
\cite{component}, fast data dissemination \cite{TON09}. In
contrast, here we consider the minimum latency broadcast
scheduling problem after channel assignment, and assume
\emph{semi-dynamic} strategy.

Collision-free minimum latency broadcast scheduling is well
studied in single-channel WANETs. Gandhi et al. \cite{UDG} show
MLBS problem in UDGs is NP-hard. Recently, Huang et al.
\cite{info09} gives an algorithm with approximation ratio of
$12$. For duty-cycle WANETs, Hong et al. \cite{ICC09} shows
MLBS to be NP-hard too, and present an algorithm with
approximation ratio $24|T|+1$ where $|T|$ is the length of one
scheduling period. Our SR-MC scenario has the multi-channel
dimension, which is not considered in single-channel and
duty-cycled WANETs. Qadir et al. \cite{MR-MC} propose several
algorithms for minimum latency broadcasting in MR-MC,
multi-rate wireless meshes. However, the proposed algorithms
depend on the multi-radio capability (i.e., multi-connection
links), and all heuristic algorithms are evaluated by
simulations without theoretical analysis. To the best of our
knowledge, \cite{ICDCN10} is the only paper to consider minimal
latency broadcast directly in multi-channel cognitive radio
networks, which is very close to us. The key difference is that
we allow channel switch while \cite{ICDCN10} assumes not.
Moreover, \cite{ICDCN10} shows the closeness of their solution
to the optimal solution through simulations. Instead, we give
two algorithms with performance guarantees.

\section{Preliminaries} \label{pre}
\subsection{Network Model and Assumptions}
The SR-MC WANETs can be modeled as a Unit Disk Graph (UDG) $G =
(V, E)$, where $V$ is the set of nodes ($|V|=n$), and $E$ is
the set of links. An edge $\{u, v\} \in E$ iff. $u$ and $v$ is
within each other's communication range. We also assume that
the time is slotted. Each time slot is equal length, and long
enough for one packet transmission and reception. Moreover, the
slot boundary is almost aligned, which can be achieved by local
synchronization protocols. We further assume that reception is
error-free if no collision happens, which is quite accurate
because control packets are often well protected by physical
layer, e.g., minimum data rate 6 Mbps in IEEE 802.11 a/g
standard. Both synchronization and error-free assumptions are
widely adopted by previous works \cite{info09, info07, ICDCN10,
MR-MC, UDG}.

The SR-MC WANETs have a total of $k$ orthogonal channels
denoted by $C=\{1, 2, ..., k\}$, and each node is equipped with
only one radio. The radio interface can be set on any channel
to transmit or listen, but not simultaneously. The reception
channel is chosen randomly from $C$ during network
initialization, which can be defined using a channel assignment
function $A$. For $v \in V$, $A(v) = c$ where $c \in C$.
Neighboring nodes may have different reception channels. In
order to enable connectivity, we assume that transmission nodes
can switch their channels to set up connections. Note
that in $G(V,E)$, the edge definition depends on topology
instead of channel since we allow channel switch. We also
assume that the neighbors' reception channels are known
beforehand, which is often achieved during neighbor discovery.


\subsection{Problem Statement}
Here we consider the single-source broadcast problem. Suppose
the source node is $s$, and the broadcast task completes when
all the other nodes receive messages sent from $s$. Assume $s$
starts the broadcast operation at time-slot $1$. Then we
formulate the MLBS problem (decision version) in SR-MC WANETs
as follows (MLBS-SRMC): \emph{Given a UDG $G(V, E)$ with
channel assignment function $A$, and positive integer $T$, is
there an assignment of time slots and transmission channels to
nodes $v \in V$, such that the broadcast scheduling is
collision-free and the schedule length is no more than $T$?}

\begin{theorem}
The MLBS-SRMC problem is NP-hard.
\end{theorem}
\begin{IEEEproof}
We prove this theorem using restriction technique \cite{NPC}.
If we restrict function $A$ to map to one single
channel, our problem is exactly the MLBS problem in
single-channel WANETs, which is NP-hard \cite{UDG}.
Hence MLBS-SRMC problem is NP-hard.
\end{IEEEproof}

Our objective can be interpreted as finding a broadcast
schedule $S = \{ S_1, ..., S_T \}$, where $S_i$, $1 \leq i \leq
T$, is the set of transmitting instances at time slot $i$, i.e.
$S_i = \{(v_0, c_0), ..., (v_j, c_j)\}, j \geq 0$. At time slot
$1$, only $s$ can transmit. After $T$ time slots, all nodes in
$V$ receive messages from $s$. Our problem can be converted to
minimize $T$. Note that if we set the cost of each edge one
unit, we can construct a Shortest-Path tree (SPT) rooted $s$.
Then the lower bound for broadcast is the depth of SPT denoted
by $l$, i.e., $T_{min} >= l$.

\subsection{Graph-Theoretic Definitions and Results}
Let $G=(V,E)$ be an undirected UDG. The subgraph of $G$ induced
by a subset $U$ of $V$ is denoted by $G[U]$. The $k$-th power
of $G$, denoted by $G^k$, is a graph over $V$ in which there is
an edge between two nodes $u$ and $v$ if and only if their
distance is $G$ is at most $k$. The minimum degree of G is
denoted by $\delta(G)$. The inductility of $G$ is defined by
$\delta^{*}(G) = \max_{U \subseteq V} \delta(G[U])$.
$\delta^{*}(G) \leq 11$ for UDG \cite{info07}. It's well-known
that the node coloring of $G$ induced by a smallest-degree-last
ordering uses at most $1 + \delta^{*}(G)$ colors
\cite{smallest-degree}. An Independent Set (IS) of a graph $G$
is a set of vertices in $G$ that no two of which are adjacent. A
maximal independent set of $G$ is not a subset of any other
$IS$ of $G$. Each node in $V$ can be adjacent to at most five
nodes in any IS of $G(V,E) $\cite{info07}, and can have at most
nineteen two-hop neighbors in any IS of $G(V,E)$ \cite{info09}.


\begin{table}
\caption{Terminology}\label{terminology}
\begin{tabularx}{8.8cm}{c|p{7cm}}
\hline
Symbol & Definition \\
\hline
 $k$& number of available orthogonal channels in $G(V,E)$\\
 $N(u)$& neighboring set of nodes $u$ in $G(V,E)$\\
 $T_{SPT}$& Shortest-Path Tree rooted $s$ in $G(V,E)$\\
 $l$& depth of $T_{SPT}$\\
 $L_i$& nodes of layer $i$ in $T_{SPT}$ \\
 $H_c$& nodes using channel $c$ in $T_{SPT}$\\
 $L_{i,c}$& nodes of layer $i$ using channel $c$ in $T_{SPT}$, $L_{i,c} = L_i \cap H_c$\\
 $M_{i,c}$& maximal independent set (dominators) of $L_{i,c}$ \\
 $P(S)$& set of parent nodes of set $S$ in $T_{SPT}$ \\
 $P_{i,c}$& parent nodes (connectors) of $M_{i,c}$ which are selected greedy \\
 $T_b$& broadcast tree constructed by Algorithm \ref{broadcast}, including nodes $V$, edges $E$ and cover function $C$ \\
 \hline
\end{tabularx}
\end{table}

\section{Basic Transmission Scheduling} \label{bts}
In this section, we give an algorithm Basic Transmission
Scheduling (BTS) for minimum latency broadcast scheduling
problem in UDGs, which is a simple extension to existing
approaches \cite{info07, ICC09}. The main notations used in
this paper are summarized in Table~\ref{terminology}.

\subsection{Algorithm Description}

Let $L_i$ be nodes of layer $i$ in $T_{SPT}$, $i = 0, 1, ...,
l$, $H_c$ be nodes using channel $c$ in $G$, $c = 1,2, ..., k$,
and $L_{i,c}$ be nodes in layer $i$ using channel $c$ ($L_{i,c}
= L_{i} \cap H_c$). Then, BTS can find a maximal independent set
$M_{i,c}$ for each $L_{i,c}$ by adding eligible nodes
sequentially. Let $P(S)$ be the set of parent nodes of $S$ in
$T_{SPT}$. Note that nodes in $P(M_{i,c})$ are not guaranteed
to be in reception channel $c$.

The key idea of BTS is to schedule collision-free
transmissions layer by layer, and channel by channel.
Take layer $i$ for example, BTS consists of two
steps:
\begin{enumerate}
\item $P(M_{i,c}) \rightarrow M_{i,c}$ sequentially for $c =
    1,...,k$;
\item $ M_{i,c} \rightarrow  L_{i,c}$ simultaneously for $c
    = 1,...,k$;
\end{enumerate}

We call $\bigcup_{c=1}^k M_{i,c}$ as layer-$i$
\emph{dominators}, and $P(\bigcup_{c=1}^k M_{i,c})$ as
layer-$i$ \emph{connectors}. For step 1), we schedule
transmissions channel by channel to avoid \emph{same-node}
collision, which means a node can be a parent of nodes in
$M_{i,c_1}$ and nodes in $M_{i, c_2}$ ($c_1 \neq c_2$). Then we
use ditance2-coloring of $P(M_{i,c})$ to achieve collision-free
scheduling to cover\footnote{In this paper, nodes are covered
means nodes receive broadcast packets.} $M_{i, c}$. Distance-2
coloring method is widely used to schedule collision-free
transmission to avoid \emph{cross-node} collision, which means, if
two nodes within two hops transmit at the same slot, there is a
collision in common neighbors. For step 2), we also schedule
collision-free transmissions in different channels
simultaneously using distance2-coloring of $M_{i,c}$. The
details are shown in Algorithm \ref{A1}. It is easy to verify
that the time complexity of BTS is $O(k n^3)$.





\begin{algorithm}[tbp] \label{A1}
\caption{Basic Transmission Scheduling}
\KwIn{$G=(V,E)$, $A$, $s$}
\KwOut{$T$, $txTime$}
$T_{SPT} \leftarrow$ SPT tree in $G$ rooted $s$; $l \leftarrow$ depth of $T_{SPT}$\;
$L_i \leftarrow$ nodes at level $i$ in $T_{SPT}$, $0 \leq i \leq l$\;
$H_c \leftarrow \{u | u \in V$ and $A(u) = c\}$, $1 \leq c \leq k$\;
\For{$i \leftarrow 1$ to $l$}
{
 \For{$c \leftarrow 1$ to $k$}
 {
    $L_{i,c} \leftarrow H_c \cap L_i$; $M_{i,c} \leftarrow \emptyset$\;
    \For{each $u \in L_{i,c}$}
    {
        \If{$N(u) \cap M_{i,c} = \emptyset$}
        {
            $M_{i,c} \leftarrow M_{i,c} \cup \{u\}$;
        }
    }
    $color_1(u) \leftarrow$ coloring by first to last ordering in $G^2[P(M_{i,c})+M_{i,c}]$ for $u \in P(M_{i,c})$\;
    $color_2(u) \leftarrow $ coloring by smallest-degree-last ordering in $G^2[L_{i,c}]$ for $u
\in M_{i,c}$\;
 }
}
$T \leftarrow 0$\;
\For{$i \leftarrow 1$ to $l$}
{
 \For{$c \leftarrow 1$ to $k$}
 {
    $txTime(u, c) \leftarrow T + color_1(u)$  for $u \in
P(M_{i,c})$\;
    $T \leftarrow T + \max_v \{color_1(v) | v \in P(M_{i,c})\}$\;
 }
 \For{$c \leftarrow 1$ to $k$}
 {
    $txTime(u, c) \leftarrow T + color_2(u)$  for $u \in M_{i,c}$\;
 }
 $T \leftarrow T + \max_{v,c} \{color_2(v) | v \in M_{i,c}\}$\;
}
return $T$, $txTime$\;
\end{algorithm}

\subsection{Performance Analysis}

Then we give a theorem that proves the correctness of BTS and
shows the upper bound of the latency given by BTS.

\begin{theorem} \label{BTS}
Algorithm BTS is correct, and provides a collision-free
broadcast scheduling with latency at most $(4k+12)*l$.
\end{theorem}
\begin{IEEEproof}
For algorithm BTS, the transmissions are scheduled layer by
layer. The transmissions in layer $i+1$ do not start until
layer $i$ ends. We consider layer $i$ ($1 \leq i \leq l$).
Assume nodes in $L_i$ are not covered, and nodes in $L_{i-1}$
are covered.  For step 1), we color $P(M_{i,c})$ in layer $i-1$
front-to-end by distance2-coloring, it is collision-free and
can cover $M_{i,c}$. Furthermore, transmissions in different
channel are sequential. Thus we can avoid \emph{same-node}
collisions. After that, nodes in $M_{i,c}$ are covered. For
step 2), because $M_{i,c}$ is the maximal independent set of
$L_{i,c}$, which is a dominating set of $L_{i,c}$, the
smallest-degree-last distance2-coloring of $M_{i,c}$ guarantees
that collision-free transmissions of $M_{i,c}$ can cover
$L_{i,c}$. Also, the parallel transmissions in different
channel are collision-free. Then all nodes in $L_i$ are
covered. Thus algorithm BTS is correct and collision-free.


We analyze the broadcast latency for layer $i$.
The cross-layer transmissions from layer $i-1$ to $i$ are channel by channel. Hence we
can consider a single channel $c$, and then multiple $k$. Note that
$M_{i,c}$ is still an independent set in UDG. Hence a node $u$ in
$P(M_{i,c})$ can have at most four neighbors in $M_{i,c}$,
because $u$ has a parent in layer $i-2$ in $T_{SPT}$, which is independent of nodes in $M_{i,c}$.
Then the distance2-coloring of $P(M_{i,c})$ uses at most four
colors. Otherwise, if a node $u \in P(M_{i, c})$ has the five
color, it means $u$ shares five neighbors with nodes in
$P(M_{i, c}) \backslash \{u\}$, i.e., connecting five neighbors
in $M_{i, c}$, which contradicts. Hence four time slots are
enough for single-channel cross-layer transmission. For
transmissions from $M_{i, c}$ to $L_{i,c}$, because $M_{i,c}$
is a maximal independent set, the smallest-degree-last ordering
distance2-coloring of $M_{i,c}$ uses at most 12 colors
\cite{info07, smallest-degree}. Since the same layer
transmission can be in parallel, we do not need to multiple
$k$. Hence, twelve time slots are enough. Given that our
analysis applies from layer $1$ to layer $l$, the overall
broadcast latency is at most $(4k+12)*l$. In other words, BTS
algorithm has approximation ratio of $4k+12$.
\end{IEEEproof}

\section{Enhanced Transmission Scheduling} \label{ets}
In this section, we present an enhanced algorithm ETS, which
has approximation ratio of $k+23$. We notice that BTS uses
sequential channel transmissions in cross-layer, which is too
conservative. Also, the strict constraint that the next layer
cannot start transmissions before last layer ends cannot
utilize the natural no-collision of multi-channel
transmissions. Based on these two observations, we propose ETS.

\subsection{Algorithm Description}
ETS is a broadcast tree based algorithm. If $u$ is the parent
of $w$, $u$ is responsible for transmitting packets to $w$
collision-free. The formal description of constructing
broadcast tree is shown in Algorithm \ref{broadcast}. As stated
in BTS, we have \emph{connectors} to connect \emph{dominators}.
ETS differs from BTS mainly in the selection of
\emph{connectors}. BTS simply selects $P(M_{i,c})$ as
\emph{connectors}, and schedules transmissions in separate
channel to avoid \emph{same-node} collision. ETS selects
\emph{connectors} greedy, i.e., selecting parent nodes to cover
maximal uncovered \emph{connectors}. Note that here we select
nodes from $L_i$.  All selected nodes to cover $M_{i+1,c}$
are recorded in $P_{i+1,c}$. For \emph{dominators} $M_{i,c}$ to
cover $L_{i,c}$, it is similar.



\begin{algorithm}[tbp]  \label{broadcast}
\caption{Broadcast Tree Construction}
\KwIn{$G(V,E), l, L_{i,c}$}
\KwOut{$T_b$}
\For{each $v \in V$}
{
    $p(v) = \emptyset$;
}
\For{$i \leftarrow 1$ to $l$}
{
 \For{$c \leftarrow 1$ to $k$}
 {
    \While{$\exists w \in L_{i,c}$ s.t. $p(w) = \emptyset$}
    {
        $u \leftarrow argmax_{u} |\{w \in N(u) \cap L_{i,c} | p(w) = \emptyset, u \in L_{i,c}\}|$\;
        $C(u) \leftarrow \{w | w \in N(u) \cap L_{i,c}$, $p(w) = \emptyset\}$\;
        \For{each $v \in C(u)$}
        {
            $p(v) \leftarrow u$\;
        }
        $M_{i,c} \leftarrow M_{i, c} \cup \{u\}$\;
    }
    \While{$\exists w \in M_{i+1,c}$ s.t. $p(w) = \emptyset$}
    {
        $u \leftarrow$ $argmax_{u} |\{w \in N(u) \cap M_{i+1,c} | p(w) = \emptyset, u \in L_i\}|$\;
        $C(u) \leftarrow \{w | w \in N(u) \cap M_{i+1,c}$, $p(w) = \emptyset\}$\;
        \For{each $v \in C(u)$}
        {
            $p(v) \leftarrow u$\;
        }
        $P_{i+1,c} \leftarrow P_{i+1, c} \cup \{u\}$\;
    }

 }
}
$V_b \leftarrow V$; $E_b \leftarrow \{(u,v)|u=p(v)\}$\;
return $T_b = (V_b, E_b, C)$\;
\end{algorithm}


The broadcast scheduling is shown in Algorithm
\ref{scheduling}. Though we still find available transmission
slot channel by channel and layer by layer, we break the
layered transmission constraint. In other words, layer $i+1$
can start before layer $i$ ends only if nodes in layer $i+1$ do
not bring collisions to already scheduling. Let step 1) be
$P(M_{i,c}) \rightarrow M_{i,c}$, and step 2) be $M_{i,c}
\rightarrow L_{i,c}$ Note that for step 1) and 2)
transmissions, we have $P_{i,c}$ and $M_{i,c}$ recorded
respectively. Hence, we first select tx node $u$ from $P_{i,c}$
($M_{i,c}$) sequentially, and then select the minimum time $t$
larger than reception time to satisfy no-collision constraints:
(1) $u$ does not bring collisions to already scheduled
transmissions in common neighbors; (2) $u$ can not transmit at
the same slot that it has been assigned to other channels. The
collision slots are recorded in $I(u)$. After scheduling all
nodes in $V$, we can find the maximal transmission time $T$.
From Algorithm \ref{broadcast} and \ref{scheduling}, we can
find that time complexity of ETS is also $O(k n^3)$.

\begin{algorithm}[tbp]
\caption{Enhanced Transmission Scheduling}
\label{scheduling}
\KwIn{$G=(V,E), T_b, l, M_{i,c}, P_{i,c}$}
\KwOut{$T$, $txTime$}
\For{each $v \in V$}
{
    $rcvTime(v) \leftarrow \infty$\;
    \For{$c \leftarrow 1$ to $k$}
    {
        $txTime(v,c) \leftarrow 0$\;
    }
}
$rcvTime(s) \leftarrow 0$\;
\For{$i \leftarrow 1$ to $l$}
{
 \For{$c \leftarrow 1$ to $k$}
 {
    \For{$j \leftarrow 1$ to $|P_{i,c}|$}
    {
        $u \leftarrow$ $j$-th node in $P_{i,c}$\;
        $I_1(u) \leftarrow \{t | \exists w \in N(u)$ that receives a message coll-free at time
$t\}$\;
        $I_2(u) \leftarrow \{t | t = txTime(u, ch) > 0, ch \neq
c\}$\;
        $txTime(u, c) \leftarrow \min\{t | t > rcvTime(u)$ and $t \not\in
I_1(u) \cup I_2(u)\}$\;
        \For{each $v \in C(u)$}
        {
            $rcvTime(v) \leftarrow txTime(u, c)$\;
        }
    }
    \For{$j \leftarrow 1$ to $|M_{i,c}|$}
    {
        $u \leftarrow$ $j$-th node in $M_{i,c}$\;
        $I_1(u) \leftarrow$ $\{t|\exists w \in N(u)$ that receives a message coll-free at time
t$\}$\;
        $txTime(u, c) \leftarrow \min\{t | t > rcvTime(u)$ and $t \not\in
I_1(u)\}$\;
        \For{each $v \in C(u)$}
        {
            $rcvTime(v) \leftarrow txTime(u, c)$\;
        }
    }
 }
}
$T \leftarrow \max_{u,c} \{txTime(u,c)\}$ for $u \in
V$, $1 \leq c \leq k$\;
return $T$, $txTime$\;
\end{algorithm}

\subsection{Performance Analysis}
We first give a lemma about the correctness of ETS.
\begin{lemma} \label{correctness}
Algorithm ETS is correct and provides a collision-free broadcast scheduling.
\end{lemma}
\begin{IEEEproof}
ETS uses no-collision rule to select transmission slot, so it
must be collision-free. We only need to prove ETS provides a
broadcast scheduling. Assume all nodes in $L_{i}$ ($1 \leq i
<l$) are covered now. We show that after transmission of
$P_{i+1,c}$ and $M_{i+1,c}$ ($1 \leq c \leq k$), all nodes in
$L_{i+1}$ are covered. For any node $u \in L_{i+1}$, $u$ must
belong to a particular set $L_{i+1,c}$. If $u \in M_{i+1,c}$,
it is covered by $P_{i+1,c}$. If $u \in L_{i+1,c}\backslash
M_{i+1,c}$, it must be covered by $M_{i+1,c}$. Because nodes in
$P_{i+1,c}$ transmit before nodes in $M_{i+1,c}$, nodes in
$L_{i+1}$ are fully covered. The proof is complete.

\end{IEEEproof}

Let $t_i$ be the time that all tx nodes in $L_i$ finish their
transmissions, $0 \leq i \leq l$. We can give following lemmas.
\begin{lemma} \label{dominators}
Let $M_i = \bigcup_{c=1}^{k} M_{i,c}$. For any $u \in M_{i}$, $\max_c \{txTime(u,c)\} \leq t_{i-1} +
20$ where $u \in M_{i,c}$, $1 \leq c \leq k$.
\end{lemma}
\begin{IEEEproof}
Note that $u$ must be in some $M_{i,c}$, since we select
$M_{i,c}$ by channel $c$. In other words, $M_{i,c_1} \cap
M_{i,c_2} = \emptyset$ for $c_1 \neq c_2$.
Given any channel $c$, for any node $u \in M_{i,c}$,
$rcvTime(u) \leq t_{i-1}$, because $p(u) \in L_{i-1}$.
Note that all interfering nodes of $u$ are also in
$M_{i,c}$. Let $N_{IS}(u,2)$ denote the set of nodes consisting of an independent set within two hops
from any node $u$. $|N_{IS}(u,2)| \leq 19$ for UDGs
\cite{info09}. In other words, for any $u \in M_{i,c}$, the maximal number of interfering
nodes is 19, because $M_{i,c}$ is an independent set of $G$.
Hence, $txTime(u,c)\} \leq t_{i-1} + 20$. Because our
argument is for any channel, this completes the proof of Lemma \ref{dominators}.
\end{IEEEproof}



\begin{lemma} \label{connectors}
Let $P_i = \bigcup_{c=1}^k P_{i,c}$. For any $u \in P_{i}$, $\max_c \{txTime(u,c)\} \leq t_{i-1} +
k + 23$ where $u \in P_{i,c}$, $1 \leq c \leq k$.
\end{lemma}
\begin{IEEEproof}
For ETS, $u$ is dominated by some node in $M_{i}$. Due to Lemma
\ref{dominators}, $rcvTime(u) \leq t_{i-1} + 20$. Let $I(u)$ be
the set of interfering slots. $|I(u)|$ must be less than the
number of interfering nodes. For $u$, the collision includes
\emph{cross-node} collision $I_1(u)$ and \emph{same-node}
collision $I_2(u)$. Assume $u \in P_{i,c}$. $u$ is responsible
for transmitting packets to $M_{i+1,c}$. Hence, $|I_1(u)| \leq
N(u) \cap M_{i+1,c}-1$. Because $u$ can connect at most five
neighbors in independent set of UDGs, and $u$ has a parent in
layer $i-1$ in $T_{SPT}$, $|I_1(u)| \leq 3$. It is trivial that
$|I_2(u)| \leq k-1$ since we only have $k$ channels. Though u
can be in $P_{i, c_1}$ ($c \neq c_1$), this constraint holds
for $u \in P_{i, c_1}$. Because $u \in P_{i,c}$ for $1 \leq c
\leq k$ selects transmission slots greedy, we only need to
consider the maximal constraint for all channels. Therefore,
$txTime(u,c) \leq t_{i-1} + 20 + (3 + k-1) + 1 = t_{i-1}+k+23$.
The proof is accomplished.
\end{IEEEproof}


Combining Lemma \ref{correctness} and
\ref{connectors}, we can have a theorem.
\begin{theorem} \label{ETS}
Algorithm ETS is correct, and provides a collision-free
broadcast scheduling with approximation ratio of $k + 23$.
\end{theorem}
\begin{IEEEproof}
For source $s$, it needs at most $k$ time slot, i.e. $t_0 \leq k$.
Nodes in $T_{i}$ transmit, and then nodes in $PN_{i}$ transmit
($1 \leq i < l$). Finally, nodes in $T_{l}$ transmit, and at
most 20 time slots are enough.
Hence, 
the overall latency is at most $k + (k+23)*(l-1) + 20 \leq (k+23)*l$. Thus,
the approximation ratio is $k+23$.
\end{IEEEproof}


\begin{figure*}
\centering
\subfigure[]{
\label{fig:subfig:a} 
\includegraphics[width=2.3in]{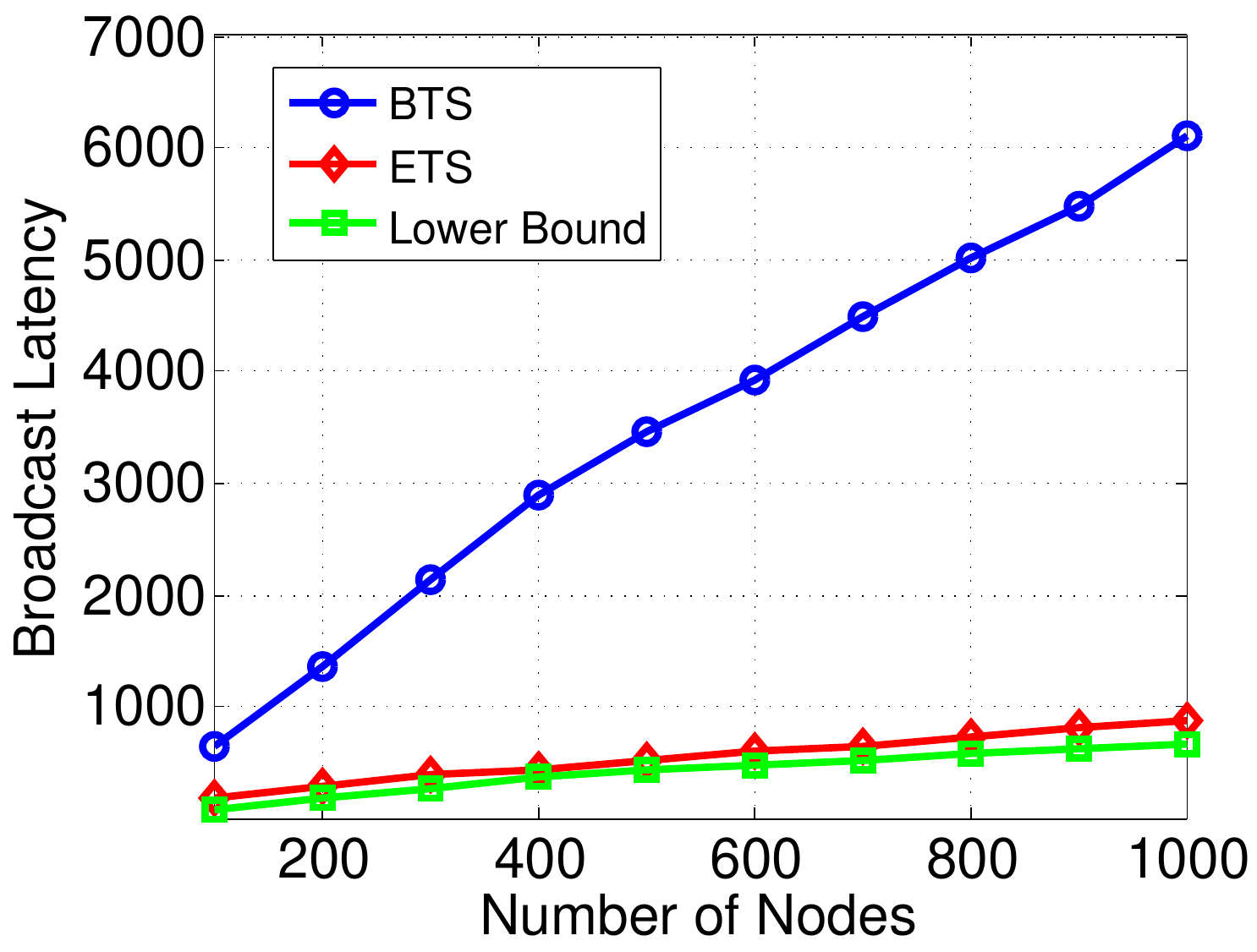}}
\subfigure[]{
\label{fig:subfig:b} 
\includegraphics[width=2.3in]{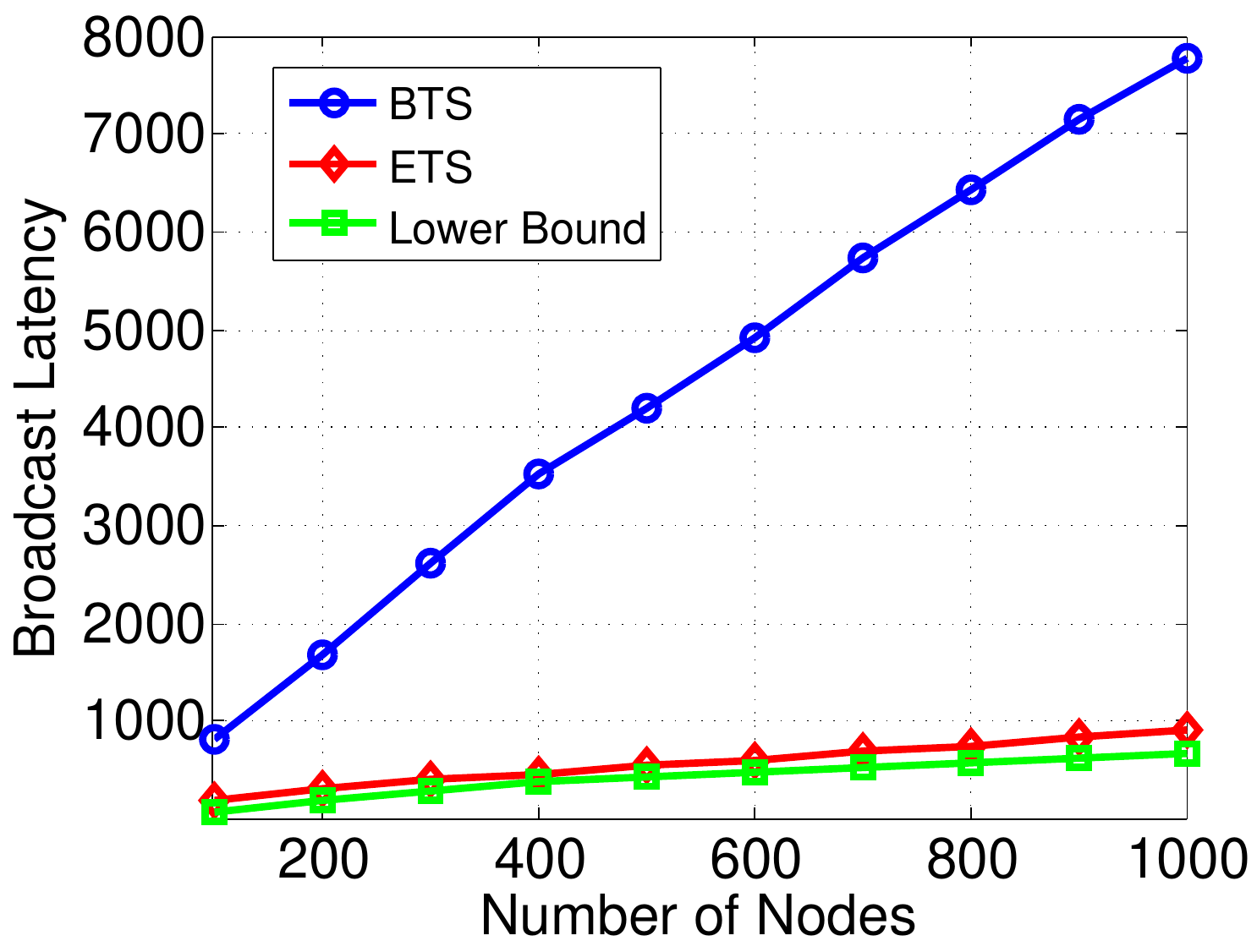}}
\subfigure[]{
\label{fig:subfig:c} 
\includegraphics[width=2.3in]{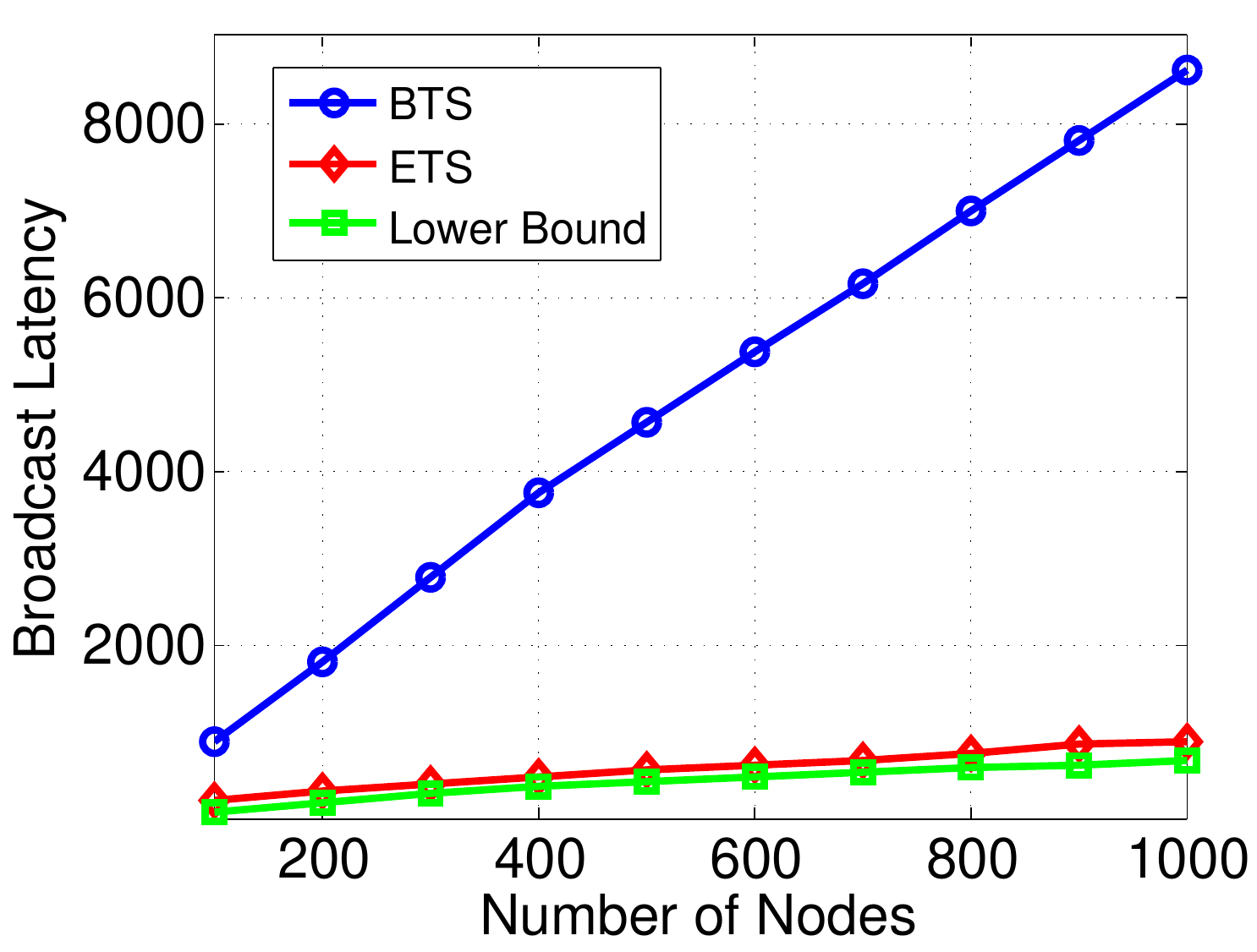}}
\caption{Broadcast latency vs. network size $n$ when (a) $k=10$; (b) $k=20$; (c) $k=30$;}
\label{fig1} 
\end{figure*}

\begin{figure*}
\centering
\subfigure[]{
\label{fig:subfig:a} 
\includegraphics[width=2.3in]{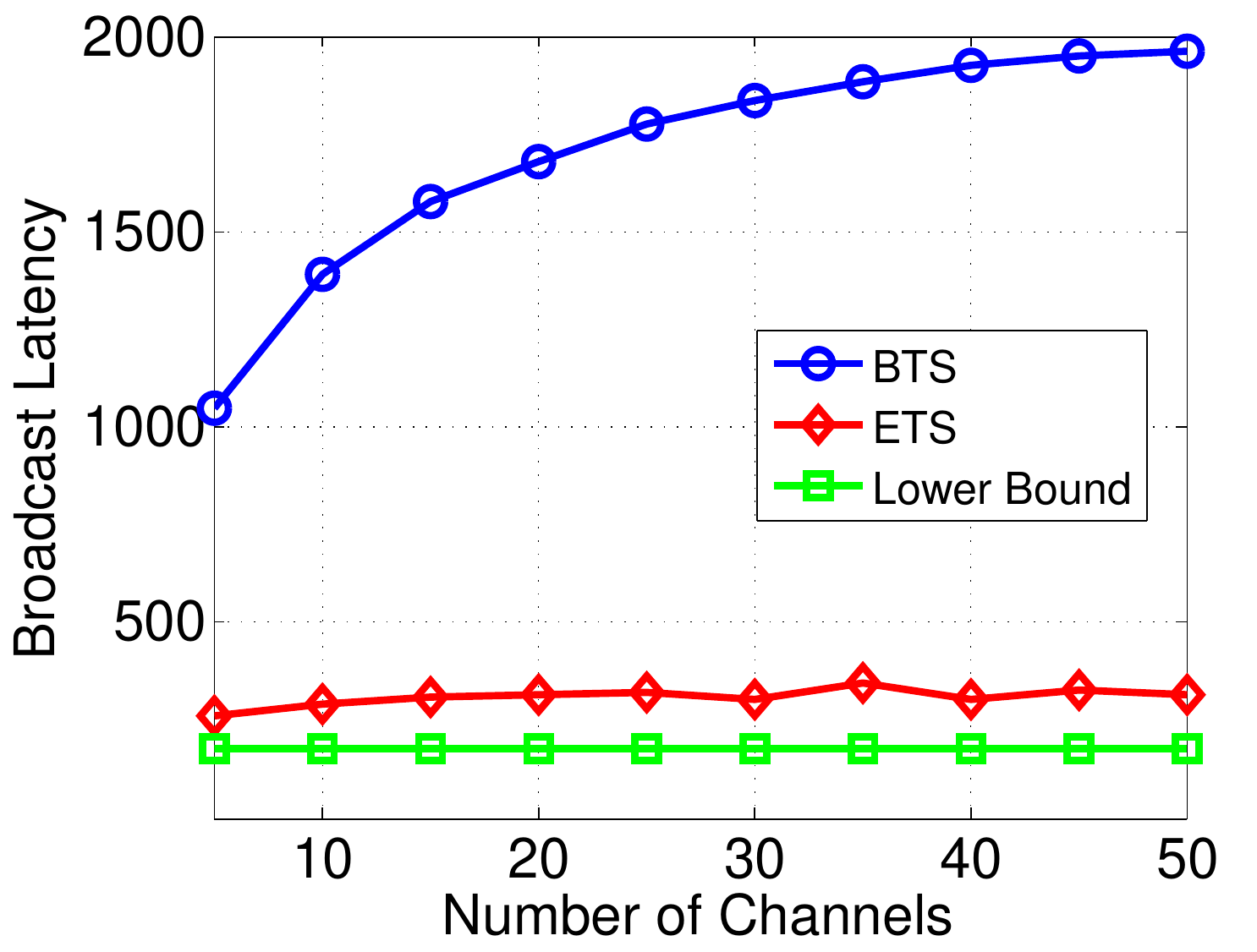}}
\subfigure[]{
\label{fig:subfig:b} 
\includegraphics[width=2.3in]{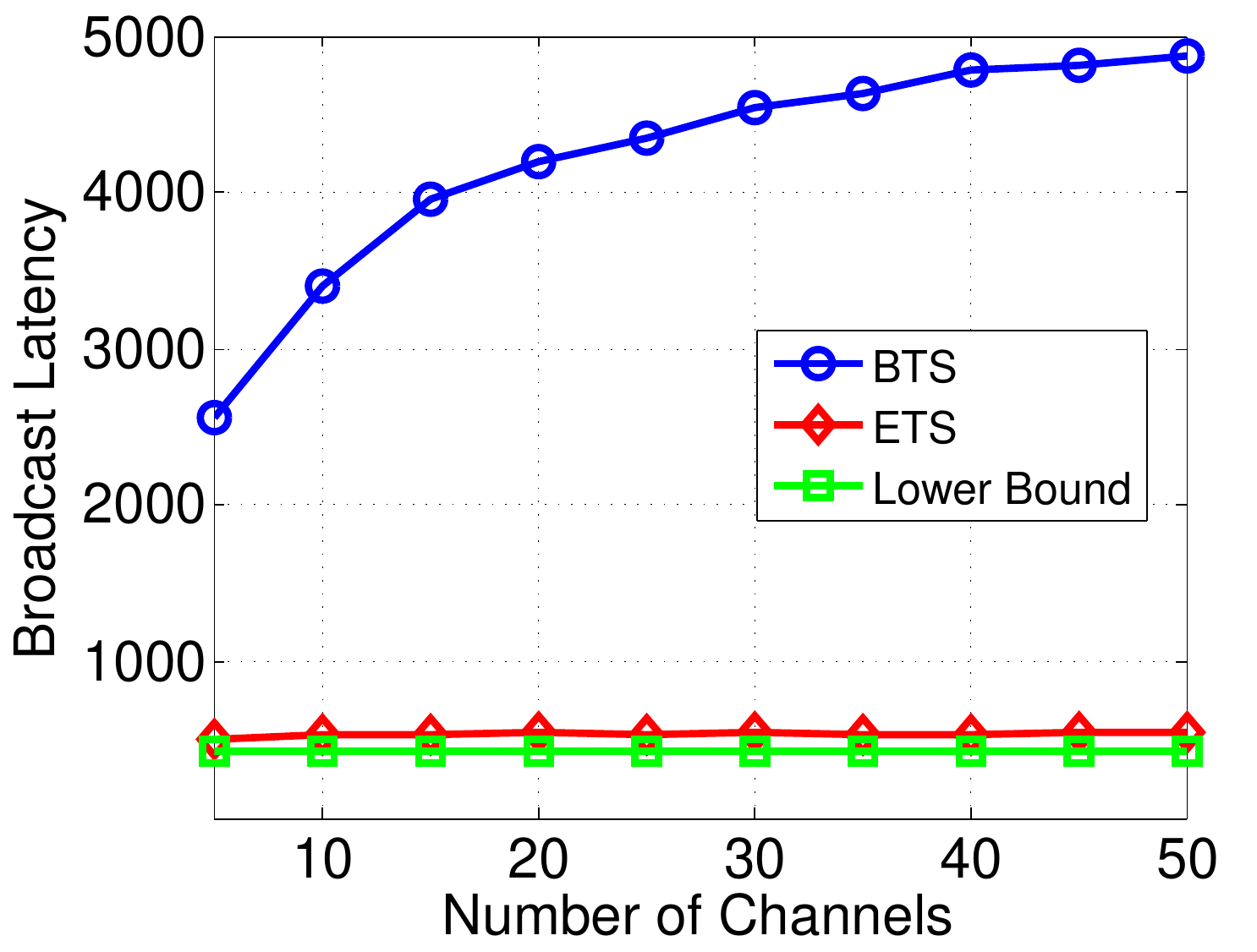}}
\subfigure[]{
\label{fig:subfig:c} 
\includegraphics[width=2.3in]{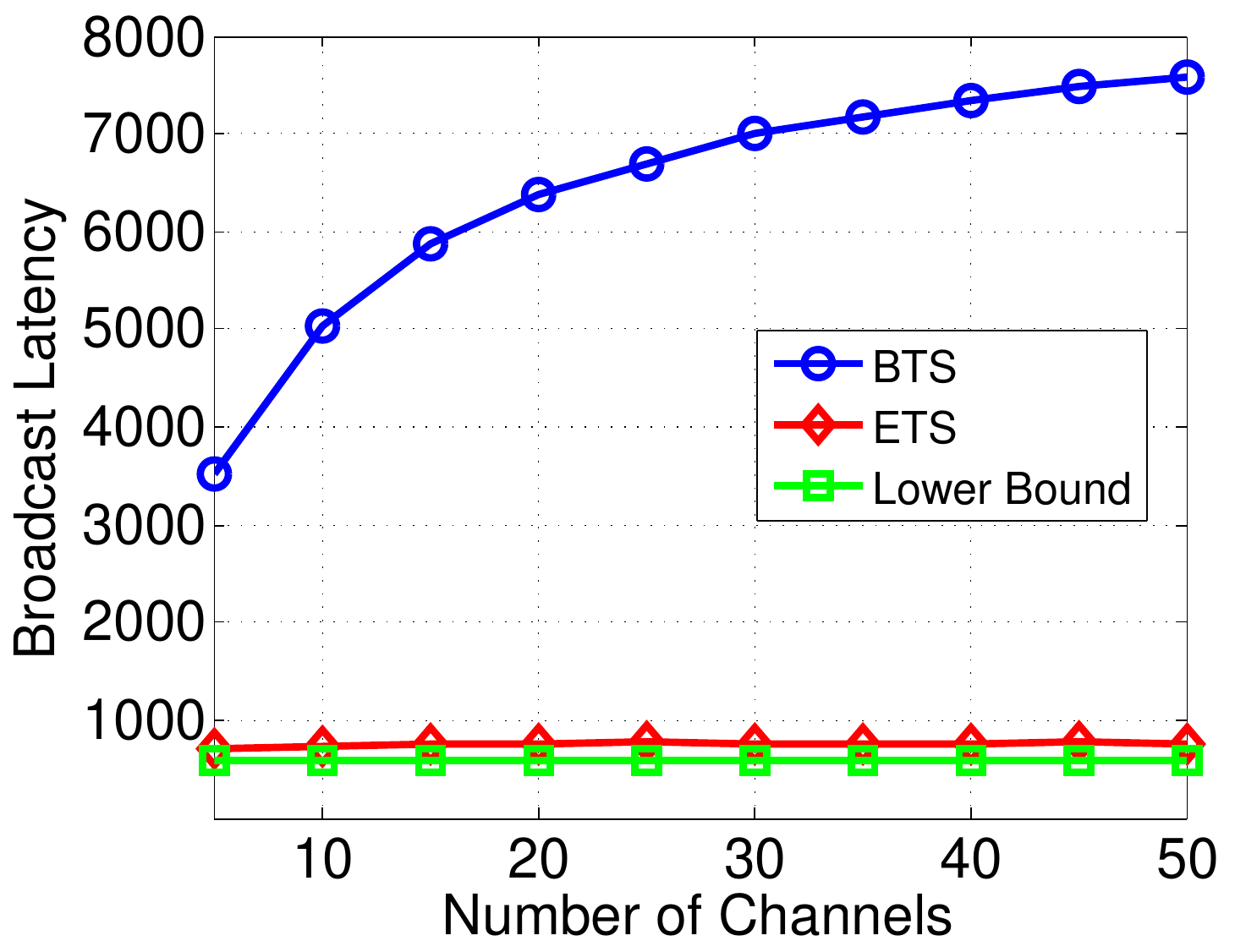}}
\caption{Broadcast latency vs. number of channels $k$ when (a) $n=200$; (b) $n=500$; (c) $n=800$;}
\label{fig2} 
\end{figure*}

\section{Performance Evaluation} \label{eval}
In this section, we run simulations to study the performance of
ETS. Since there is no directly applicable algorithms in SR-MC
WANETs, we use BTS as benchmark. The metric is broadcast
latency. To show the optimal result, we also plot the lower
bound of broadcast latency using the depth of $T_{SPT}$. We
consider the impact of number of nodes $n$ and number of
orthogonal channels $k$. To increase the depth of constructed
$T_{SPT}$, we vary the network area with respect to $n$. For
example, when $n=1000$, the area size is $1000 \times 1000$
$m^2$. All nodes are randomly deployed in corresponding areas,
and their reception channels are randomly choosen from $C$. We
run the simulation 10 times, and show the average results. For
each time, we generate a new topology and channel assignment.





First we evaluate the impact of $n$, which ranges from $100$ to
$1000$ with step $100$. Simulation results with different $k =
10, 20, 30$ are shown in Figure \ref{fig1}. It is obvious that
ETS performs better than BTS. More importantly, the performance
of ETS is close to the lower bound, which demonstrates the gain
of parallel multi-channel transmissions. Furthermore, when $k$
becomes larger, the broadcast latency also raises due to the
more nodes in each channel set, but the linear tendency keeps.
Note that, for approximation ratio, the performance of both
algorithms are much smaller than theoretical results. It can be
explained that our theoretical analysis considers the worst
case, but probably we are not in the worst case. 

Then we study the impact of $k$, which is set from 5 to 30 with
step 5. Simulation results with $n = 200, 500, 800$ are shown
in Figure \ref{fig2}. The performance of ETS is still better
than BTS, and close to the lower bound due to the same reason
mentioned above. Note that the scale of y-axis in Figure 2(a)
is smaller. Here, the depth of $T_{SPT}$ keeps almost constant
since $n$ does not change (i.e., network size does not change).
However, for BTS, the broadcast latency grows sub-linearly with
respect to $k$. Intuitively, with larger $k$ and same $n$,
though the number of sequential transmissions raise linearly
with respect to $k$, the transmissions in each channel reduce
since nodes in each channel are less. For Figure \ref{fig1}
with larger $n$ and constant $k$, transmissions in
single-channel increase, but the number of sequential channel
transmissions is the same. It can explain the linear and
sub-linear phenomenon in Figure \ref{fig1} and \ref{fig2}.

\section{Conclusion} \label{con}
In this paper, we consider the minimum latency broadcast
scheduling problem in SR-MC WANETs. We first identify the
challenge and opportunity in such networks. To solve the
NP-hard problem, we give an algorithm BTS with approximation
ratio of $4k + 12$, which is modified from classical
algorithms. Then we propose an algorithm ETS with approximation
ratio of and $k+23$. Both have time complexity $O(k n^3)$. The
simulation results show ETS improves the performance over BTS
significantly, and come close to the lower bound. In the
future, we want to complete our works by considering
distributed algorithms, and finding algorithms for broadcast
scheduling under \emph{dynamic} channel assignment strategy.

\end{document}